\documentclass[11pt]{article}
\usepackage{graphics,epsfig}
\usepackage[latin1]{inputenc}
\usepackage[english,activeacute]{babel}
\usepackage[]{graphicx}
\usepackage{amsmath, amsthm, amsfonts, amssymb}
\usepackage{amssymb}
\usepackage{verbatim}

\newcommand{\nd}{\noindent}

\begin{document}

\newtheorem{theo}{Theorem}[section]
\newtheorem{definition}[theo]{Definition}
\newtheorem{lem}[theo]{Lemma}
\newtheorem{prop}[theo]{Proposition}
\newtheorem{coro}[theo]{Corollary}
\newtheorem{exam}[theo]{Example}
\newtheorem{rema}[theo]{Remark}
\newtheorem{example}[theo]{Example}
\newtheorem{principle}[theo]{Principle}
\newcommand{\ninv}{\mathord{\sim}}
\newtheorem{axiom}[theo]{Axiom}

\title{A formal framework for the study of the
notion of undefined particle number in quantum
mechanics}

\author{{\sc Newton C.A. da Costa}$^{1}$ \ \ {\sc
 and} \ \ {\sc Federico Holik}$^{2,\,3}$}

\maketitle

\begin{center}

\begin{small}
1- Department of Philosophy, Federal University of Santa Catarina, Florian\'{o}polis, SC 88040-900, Brazil \\
2- Instituto de F\'{i}sica La Plata (IFLP), CONICET, 115 y 49, 1900 La Plata, Argentina\\
3- Center Leo Apostel for Interdisciplinary Studies and, Department of Mathematics, Brussels Free University Krijgskundestraat 33, 1160 Brussels, Belgium \\
\end{small}
\end{center}

\maketitle

\vspace{1cm}

\begin{abstract}
\noindent It is usually stated that quantum mechanics presents problems with the identity of particles, the most
radical position -supported by E. Schr\"{o}dinger- asserting that
\emph{elementary particles are not individuals}. But the subject
goes deeper, and it is even possible to obtain states with an
undefined particle number. In this work we present a set theoretical
framework for the description of undefined particle number states in quantum mechanics which provides a precise
logical meaning for this notion.
This construction goes in the line of solving a problem posed by Y. Manin, namely,
\emph{to incorporate quantum mechanical notions at the foundations of mathematics}.
We also show that our system is capable of representing quantum superpositions.
\end{abstract}
\bigskip
\noindent

\begin{small}
\centerline{\em Key words: set theory-undefined
particle number}
\end{small}

\bibliography{pom}

\section{Introduction}\label{s:Introduction}

Quantum mechanics (QM) in both of its versions, relativistic and
non-relativistic, is considered as one of the most important
physical theories of our time, giving rise to spectacular
technological developments and experimental predictions. Yet,
interpretation of QM still gives rise to difficult problems, which
are far from finding a definitive solution. This is, perhaps, one of
the most interesting features of QM, and poses important
philosophical questions. In particular, while classical extensional
mereology is widely investigated in important philosophical
textbooks (see for example \cite{Simons} for a complete study), the
development of a quantum mereology (i.e., a mereology based on
objects obeying the laws of QM) is still lacking. And this is an
important issue for ontological considerations, because it is
expected that a quantum mereology will be quite different than
classical extensional mereology (at least, if we follow the standard
interpretation of QM and many other interpretations as well).

The development of formal systems in which mereological properties
(or features) of a given ontology are rigorously expressed is a
helpful goal. This is the case in Le\'{s}niewski's Mereology (based
on his ``Calculus of Names") or the ``Calculus of Individuals'' of
Leonard and Goodman \cite{Simons}.

\vskip6mm

In this work we will develop a formal framework which captures
important features of the quantum formalism, namely,

\begin{itemize}

\item \emph{undefined particle number} and

\item \emph{undefined properties} (as the ones appearing in quantum superpositions).

\end{itemize}

\noindent By capturing these quantum features, our system may be
helpful for the task of developing a quantum mereology in a rigorous
way. This is an important issue for any philosopher interested in
the development of an ontology based on quantum mechanics.

\vskip3mm

We will present a construction which goes in the direction of
solving a problem posed by Y. Manin (see \cite{fyk} for a complete
discussion of Manin's problem and an alternative proposal of
solution for it). In his words \cite{Manin,Manin2}

\begin{quote}``We should consider the possibilities of developing a totally new
language to speak about infinity. Set theory is also known as the
theory of the `infinite'. Classical cri\-tics of Can\-tor (Brou\-wer
\textit{et al.}) argued that, say, the general choice axiom is an
illicit extrapolation of the finite case.

I would like to point out that this is rather an
extrapolation of common-place physics, where we can distinguish
things, count them, put them in some order, etc. New quantum physics
has shown us models of entities with quite different behavior. Even
`sets' of photons in a looking-glass box, or of electrons in a nickel
piece are much less Cantorian than the `set' of grains of sand. In
general, a highly probabilistic `physical infinity' looks
considerably more complicated and interesting than a plain infinity of `things'.''\cite{Manin}
\end{quote}

\noindent Thus, Manin suggests the development of set theories
\cite{Manin-SetTheory,Halmos,Kunen,BrignoleDaCosta} incorporating
the novel features of quantum entities, which depart radically from
our every day concepts\footnote{Although Manin has seemingly changed
his position regarding this subject \cite{Manin-SetTheory}, the
problem posed above still seems interesting to us and we will take
it as a basis for our work.}. In this line, many alternatives where
developed, most of them grounded in non-reflexive logics
\cite{daCosta1980,daCosta2009}. In particular, it is possible to
incorporate in a Zermelo-Frenkel ($ZF$) set theory the notion of
quantum non-individuality \cite{fyk,Dalla Chiara,una compara,Un
estudio,Why quasisets} and this was done by introducing
indistinguishability ``right at the start" \cite{Heinz Post}.
According to the interpretation of E. Schr\"{o}dinger an elementary
particle cannot be considered as an individual entity

\begin{quotation}``I mean this: that the elementary particle is not an individual; it
cannot be identified, it lacks `sameness'. The fact is known to
every physicist, but is rarely given any prominence in surveys
readable by nonspecialists. In technical language it is covered by
saying that the particles `obey' a newfangled statistics, either
Einstein-Bose or Fermi-Dirac statistics. [...] The implication, far
from obvious, is that the unsuspected epithet `this' is not quite
properly applicable to, say, an electron, except with caution, in a
restricted sense, and sometimes not at all." E. Schr\"{o}dinger
(\cite{Sch98}, p.197)\end{quotation}

\noindent Similarly, Michael Redhead and Paul Teller claim in
\cite{redtel91,redtel92} that:

\begin{quotation}``Interpreters of quantum mechanics largely agree that
classical concepts do not apply without alteration or restriction to
quantum objects. In Bohr's formulation this means that one cannot
simultaneously apply complementary concepts, such as position and
momentum, without restriction. In particular, this means that one
cannot attribute classical, well defined trajectories to quantum
systems. But in a more fundamental respect it would seem that
physicists, including Bohr, continue to think of quantum objects
classically as individual things, capable, at least conceptually, of
bearing labels. It is this presumption and its implications which we
need to understand and critically examine.'' M. Redhead and P.
Teller (\cite{redtel92}, p.202)\end{quotation}

\noindent It is important to mention that, besides the conception of
quantum entities as non-individuals, the validity of the principle
of identity of indiscernibles (PII) in QM was also questioned (see
for example \cite{fyk}, \cite{FrenchRedheadPIIviolation} and
\cite{ButterfieldPIIviolation}). PII can be written as follows: it
is not possible for two \emph{individuals} to possess all the same
attributes in common \cite{FrenchRedheadPIIviolation}. As remarked
in \cite{FrenchRedheadPIIviolation}, if quanta were not individuals,
``PII would not be either true or false, but simply inapplicable".
Thus, violation of PII and non-individuality of quanta are not
equivalent and should not be confused.

In the last years, a different perspective on the problem of quantum
indistinguishability was developed \cite{Saunders2003,Saunders2006}.
In \cite{MulSaun-2008,Muller-2009} (see also \cite{Riseoftheappes})
it is claimed that according to quantum theory, indistinguishable
particles are not utterly indiscernible, but obey a weaker form of
discernibility, namely, \emph{weak discernibility}. This weak form
of discernment is achieved by a relational symmetric and non
reflexive relation between the relata. Different grades of
discernibility in standard model theory and its logical relations
and links with philosophical problems are discussed in
\cite{Caulton-Butterfield-2} and \cite{Ladyman-Linnebo-Pettigrew}
(see also \cite{Caulton-Butterfield-1}).

Though these works are very compelling, the success in their
application to the problem of distinguishability of elementary
particles is far from being conclusive. In the first place, the
results presented in \cite{MulSaun-2008,Muller-2009} were criticized
in \cite{Caulton-DiscerningIndiscernible}\footnote{In the Concluding
remarks of \cite{Caulton-DiscerningIndiscernible}, Caulton claims
that the approaches of Muller, Saunders and Seevinck ``...have been
seen to fail, due to their surreptitious use of mathematical
predicates that can be given no physical interpretation."}, because
the properties used to discern (weakly) were unphysical, a
perspective to which we adhere\footnote{Similarly, in
\cite{DieksPII} it is claimed that (our emphasis): ``All evidence
points into the same direction: `identical quantum particles' behave
like money units in a bank account rather than like Blackean
spheres. It does not matter what external standards we introduce,
they will always possess the same relations to all (hypothetically
present) entities. The irreflexive relations used by Saunders and
others to argue that identical quantum particles are weakly
discernible individuals \emph{lack the physical significance
required to make them suitable for the job}."}. But the solution
proposed in \cite{Caulton-DiscerningIndiscernible} is not very
attractive either: particles are weakly discerned by using an
observable based on their (squared) relative positions in space. But
one may wonder how is it possible to discern something in this way,
given that it is widely known how difficult is to assign definite
positions to particles previous to any measurement. It seems that
the only thing achieved here is numerical distinctness of space-time
points, something which in principle should not be equated with
discernibility of the particles involved (unless cumbersome
interpretational moves are made)\footnote{Related to this
observation, see also \cite{DieksPII} where a similar argument can
be found for spins and the following observation is made regarding
position measurements in QM: ``To see how this complicates matters,
think of a one-particle position measurement carried out on a
many-particles system described by such a symmetrized state. The
result found in such a measurement (for example, the click of a
Geiger counter or a black spot on a photographic plate) is not
linked to one of the `particle labels'; it is, in symmetrical
fashion, linked to all of them. This already demonstrates how the
classical limit of quantum mechanics does not simply connect the
classical particle concept to individual indices in the quantum
formalism".}. A similar observation applies to observables different
than position. Indeed, a similar problem seems to appear in
\cite{Riseoftheappes}, where the case of two entangled bosons is
discussed.

It may be argued that the relation of weak discernibility holding
between two electrons, can probably ensure that the number of
objects is indeed two. But it seems that it falls short of
separating them in such a way that they can be successfully
identified. Indeed, in \cite{Bigaj} it is pointed out that:

\begin{quotation}
``\noindent One sense of discerning involves recognizing some
qualitative differences (whether in the form of different properties
or different relations) between the objects considered. When we
discern objects in this sense, we should (at least in principle) be
able to pick out one of them but not the other. Being able to
discern objects in that way seems to be a prerequisite for making
successful reference, or giving a unique name, to each individual
object. But by discerning we can also mean recognizing objects as
numerically distinct. In this sense of the word, discernment is a
process by which, using some qualitative features of the objects, we
make sure that there are indeed two entities and not one."
\end{quotation}

\noindent In \cite{Ladyman-Bigaj}, the notion of
witness-discernibility is used to argue against the use of weak
discernibility as a means to rehabilitate PII in QM. Even the very
applicability of the notion of weak discernibility in the quantum
framework was criticized in \cite{DieksPII}. Taking into account the
different criticisms mentioned above, the conclusion that weak
discernibility entails a recovering of PII and discernibility of
quanta is too hasty. Put in the words of Dieks \cite{DieksPII}:

\begin{quotation}
``The analogy between quantum mechanical systems of ``identical
particles'' and classical collections of weakly discernible objects
is only superficial. There is no sign within standard quantum
mechanics that ``identical particles'' are things at all: there is
no ground for the supposition that relations between the indices in
the formalism possess physical significance in the sense that they
connect actual objects. Consequently, the irreflexivity of these
relations is not important either. Conventional wisdom appears to
have it right after all.''\cite{DieksPII}
\end{quotation}

\noindent So, even if the approach based on weak discernibility
could be developed in the future in order to provide a more
attractive solution to the problem of discerning elementary
particles, none of the results presented up to now is conclusive.
The plausibility of non-individuals was defended in
\cite{JonasWitherAway2013,JonasDecio}, the validity of PII was
questioned in
\cite{JonasPII,FrenchRedheadPIIviolation,ButterfieldPIIviolation}
and different criticisms against weak discernibility are presented
in
\cite{Bigaj,Ladyman-Bigaj,DieksPII,Hawley2006,Hawley2009,vanFraassenPII2008}.
Furthermore, even from the perspective of weak discernibility
approach, quanta are not individuals in the sense that they cannot
be absolutely discerned by qualitative physical properties. In this
way, the question regarding individuality or non-individuality of
quanta remains unsettled.

\vskip3mm

In a similar vein, the usual assumption that a \emph{definite
particle number} can be always obtained was also criticized. This
conclusion is grounded in the well known result that it is not
possible to assign in general, previous to measurement, definite
values to observables in superposition states \cite{Mittelstaedt}.
Thus, a new turn of the Manin's problem was presented in
\cite{Holik-Thesis,Holik-Thesis2,Domenech-Holik,
Holik-NoNameNonumber,Domenech-Holik-deRonde,Domenech-Holik-Krause}.

In this work we will follow the interpretation of QM which denies
that quantum systems can be always considered as singular unities (a
quantum system as a ``one"), or collections of them (a quantum
system as a ``many").

It is important to remark here that there are other interpretations
which deny the existence of systems with undefined particle number.
In such interpretations, states which involve superpositions with
different particle number are usually interpreted as ordinary
mixtures. Another possibility may be to consider P. Teller's notion
of non-supervenient relations in order to describe superpositions in
particle number. Regarding this last possibility, we quote Teller
\cite{Teller-Supervenient-1}:

\begin{quotation}
``Supervenience provides an attractive answer to this question,
attractive because the answer is consistent with the absence of
explicit reductions or definitions of the non-physical in terms of
the physical. For example, a physicalist might claim that mental
states supervene on brain or other bodily states, in the sense that
two physically identical bodily states would exhibit the same mental
states, even though these mental states might well not be definable
in terms of the bodily states."
\end{quotation}

\noindent We see that for Teller, the `attractiveness' of the
approach based on supervenience lies in the fact that there can be
no explicit reductions or definitions in terms of the relata. But
from the point of view of \emph{relational
holism}\cite{Teller-Supervenient-1}\footnote{See also
\cite{Teller-Supervenient-2} and \cite{MorgantiRelationalHolism} for
a development of this notion and the problems posed by Teller.}, it
is plausible that there exist collections of objects having physical
relations which do not supervene on the non-relational physical
properties of the parts. This would be the case for entangled states
in QM, such as those violating Bell's inequalities. By continuing
this, one may try to explain states with undefined particle number
as a kind non-supervenient relation between the particles involved
in the terms of the superposition. But undefined particle number
should not be confused with entanglement; it is an undefined
property of the system \emph{as a whole}: the superposition
describes a state of affairs in which one of the properties of
\emph{the whole collection} is \emph{undefined}, in this case,
particle number. While undefined particle number states may present
non-local correlations (i.e., they violate some kind of Bell
inequality), these two effects should not be confused. This
distinction suggests that undefined particle number could not be
described as a non-supervenient relational property between the
particles involved, simply because it is not a well defined property
at all. These considerations are very probably not sufficient to
rule out a description of undefined particle number as a
non-supervenient relation, but this is not determinant for our
concerns in this article.

Our interest in this work is not to settle the question about which
is the correct interpretation. We focus on the development of a
framework for studying the consequences of assuming that undefined
particle number states actually exist. Notwithstanding, it is very
important to remark here that the formal framework presented in this
work contains a copy of the standard approach to mathematics (see
Section \ref{s:SetTheoryAxioms}). This implies that \emph{any
mereological construction which can be attained in a standard set
theoretical framework can also be attained with ours}. Thus, our
mereological framework \emph{has the advantage of being able to cope
with different interpretations of quantum phenomena}. In particular,
the approaches of Muller and Saunders
\cite{MulSaun-2008,Muller-2009} or a possible description of
superpositions in particle number in terms of non-supervenient
relations (in case they can be accommodated within standard formal
frameworks) can be perfectly described in our framework.

\vskip3mm

\emph{The considerations mentioned above point in the direction that
a non-standard mereology is worth to be developed}. Firstly, because
metaphysical underdetermination does not single out a unique
interpretation for quantum theory, and as we have mentioned above,
the different alternatives remain inconclusive. In particular, the
standard interpretation of QM ---asserting that superpositions
represent states of affairs in which no definite values can be
assigned to the superposed property--- remains strong. Secondly,
because in order to discuss about different interpretations, it is
important to have at hand formal frameworks in order to cope with
them, trying to capture (or to describe) in a precise (rigorous) way
the essence of the intuitive notions involved. Thus, we present here
a mereological framework powerful enough to describe different
interpretations of the quantum formalism oriented to the problem of
undefined particle number.

\vskip3mm

We will face the problems linked to \emph{undefined particle number}
and -going in the line of the Manin's problem- we will develop a
formal set theoretical framework capable of incorporating such a
quantum mechanical feature. We will also see that our framework is
capable of describing undefined properties arising from quantum
superpositions. This is the reason why our system could be
considered as a solution to a generalization of the problem possed
by Manin (see also \cite{fyk} for an alternative solution
considering non-individuality of quanta). We believe that the formal
setting presented in this work could be a concrete step
---for interpretational purposes--- to give a precise logical
meaning to what is meant by ``undefined particle number" by
incorporating this notion into a set theoretical framework. And also
that it constitutes in itself an interesting structure for the
possible development of new non-standard mathematics, which in turn,
could be the basis for new formal frameworks with potential
applications to physics. As an example of this procedure see
\cite{Domenech-Holik-Krause}. At the same time, the developments
presented in this work constitute a concrete step in order to
develop a quantum mereology.

\vskip3mm

Before entering into the content of the article, it is important to
mention that there is another important branch of formal
developments induced by quantum mechanics, namely, a vast family of
quantum logics. Since the seminal paper of Birkhoff and von Newmann
\cite{BvN}, several investigations were motivated in the fields of
logic, algebraic logic, and the foundations of physics. Besides
these developments, some authors have claimed that according to the
logical structure of QM, we should \emph{abandon classical logic}
(see for example \cite{Putnam}). On the other hand, the nowadays
dominant interpretation of the quantum logical formalism developed
by Birkhoff and von Neumann considers it as the study of algebraic
structures linked to QM, and by no means is considered as an
alternative to classical logic. Notwithstanding, it is important to
remark that there are several examples of modifications of classical
logic in the following sense. Even if it is a subtle matter to
define exactly what classical logic is, it is possible to consider
it as having two levels:

\begin{itemize}
\item 1) an \emph{elementary level}, which is
essentially first order predicate calculus, with or without
identity, and

\item 2) a \emph{non elementary level}, which
could be a set theory, a category theory, or a theory of logical
types.

\end{itemize}

It is then possible to modify level $2$ in order to develop a family
of logics which can be considered non-classical. Indeed, the system
presented in this paper in non-classical in the sense mentioned
above. It is also possible to modify level $1$, as shown in
\cite{Pavicik}. Of course, the existence of these possibilities does
not suffices to settle the question about the adequacy or non
adequacy of classical logic. Thought we will not discuss this
subject in detail in this paper, we remark that it is a matter of
fact that the influence of $QM$ in the development of formal systems
gave rise to a considerable proliferation of investigations
\cite{mackey57,jauch,piron,kalm83,kalm86,vadar68,vadar70,greechie81,gudderlibro78,giunt91,pp91,belcas81,dallachiaragiuntinilibro,
dvupulmlibro,HandbookofQL,aertsdaub1,
aertsdaub2,FR81,extendedql,Holik-Massri-Ciancaglini-2010}, including
the development of ``quantum set theories'' \cite{Takeuti,Osawa1}.

\vskip3mm

The article is organized as follows. In section
\ref{s:UndefinedNumberOverview}, we discuss the meaning of
superpositions of particle number eigenstates in Fock-space,
introducing the interpretation which supports the existence of
undefined particle number states\footnote{The Fock-space formulation
is also discussed with great detail in \cite{fyk}, Chapter $9$. See
also \cite{Domenech-Holik-Krause} and \cite{Holik-Kniznik}.}. In
section \ref{s:preliminaries} we present the preliminary notions of
our set theoretical framework by introducing its specific axioms.
After doing this, we are ready to show how our framework solves the
problem of incorporating undefined particle number in section
\ref{s:AsolutionToTheProblem}, and also that it is capable of
describing quantum superpositions. We will also present in this
Section some special features of our axiomatic and general remarks
about our construction, which could be useful for further
developments. Finally, we pose our conclusions in
\ref{s:FinalConclusions}.

\section{Undefined particle number: an
overview}\label{s:UndefinedNumberOverview}

QFT requires an understanding of states with no definite particle
number and, as explained above, we shall attempt to construct a
formal framework accommodating that notion. In order that a
superposition of states with different particle number occur, it is
necessary to have a space which includes states with different
particle number. This is provided by the Fock-Space formalism
($FSF$). The $FSF$ is used, for example, in the \emph{second
quantization formalism}, and we find a version of it both in
relativistic and non-relativistic quantum mechanics. It can be shown
that the $FSF$ may be used as an alternative approach to non
relativistic quantum mechanics \cite{Robertson}. This can be seen by
using the heuristic approach presented in elementary expositions
like \cite{Ballentine-Libro,Robertson} (but see for example
\cite{Clifton-Halvoroson-2001}, \cite{bratelli} and \cite{delaHarpe}
for a mathematically rigorous presentation). For an important
introduction to the philosophical problems of quantum field theory
(in which the FSF and particle number superpositions are discussed)
we refer to \cite{Huggett-QFT}.

We will concentrate here on coherent states of the electromagnetic
field in order to make the exposition simpler. But it is important
to remark that there are other more involved examples of
undetermined particle number, as is the case of Rindler quanta
\cite{Clifton-Halvoroson-2001} or the BCS state of Bose-Einstein
condensates \cite{Ballentine-Libro}, but we will not treat them
here.

The second quantization approach to QM has its roots in considering
the Schr\"{o}dinger's equation as a \emph{classical field equation},
and its solution $\Psi_{n}(\mathbf{r}_{1},\ldots,\mathbf{r}_{n})$ as
a \emph{classical field to be quantized}. This alternative view was
originally adopted by P. Jordan \cite{Jordan-1,Jordan-2}, one of the
foundation fathers of quantum mechanics, and spread worldwide after
the Dirac's paper \cite{Dirac}. And it is a standard way of dealing
with relativistic quantum mechanics (canonical quantization). The
space in which these quantized fields operate is the Fock-space.

It is important to remark that the $n$ particle Schr\"{o}dinger wave
equation is not completely equivalent to its analogue in the
Fock-space formalism. Only solutions of the Fock-space equation
which are eigenvectors of the particle number operator with particle
number $n$ can be solutions of the corresponding $n$ particle
Schr\"{o}dinger wave equation. And the other way around, not all the
solutions of the $n$ particle Schr\"{o}dinger wave equation can be
solutions of the Fock equation, only those which are symmetrized do.
\emph{Then, both conditions, definite particle number and
symmetrization, must hold in order that both formalisms yield
equivalent predictions.}

The hamiltonian of the $m$th mode of a quantized electromagnetic
field can be written in terms of the creation and annihilation
operators $a_{k}^{\dag}$ and $a_{k}$ as follows

\begin{equation}
H_{n}=\hbar\omega(a_{k}^{\dag}a_{k}+\frac{1}{2})
\end{equation}

\nd and so, each $a_{m}^{\dag}$ ($a_{m}$) creates (annihilates) a
photon in mode $m$. Then, a fock space state (with definite particle
number) can be expressed as

\begin{equation}\label{e:StatePhotons}
|n_{1},n_{2},\ldots,n_{m},\ldots\rangle=|n_{1}
\rangle\otimes|n_{2}
\rangle\otimes\ldots\otimes|n_{m}
\rangle\otimes\ldots
\end{equation}

\nd with $n_i$ the number of photons present in each mode of the
field. If for simplicity we concentrate in only one frequency mode
of the field, we can create any normalized superposition of states,
and in particular, the famous \emph{coherent state}

\begin{equation}\label{e:CoherentState}
|z\rangle=
\exp(-\frac{1}{2}|z|^{2})\sum_{n=0}^{\infty}\frac{
z^{n}}{(n!)^{\frac{1}{2}}} |n\rangle
\end{equation}

\nd which can be realized in laboratory \cite{Ballentine-Libro}.
State \eqref{e:CoherentState} is clearly a superposition of
different photon number states and thus is not an eigenstate of the
particle number operator. It follows that, according to the standard
interpretation, it represents a physical system formed by an
undefined number of photons. It is important to remark that there
are -at least- two interpretations of \eqref{e:CoherentState}

\begin{itemize}

\item $1$-Equation (\ref{e:CoherentState})
represents an statistical mixture of states with definite particle
number.

\item $2$-Equation (\ref{e:CoherentState})
represents an state which has no definite particle number.

\end{itemize}

\nd The orthodox interpretation of QM points in the direction of the
second option and the first one is very difficult to sustain unless
involved hypotheses are made \cite{Mittelstaedt}. Regardless the
interpretational debate, it will suffice for us that \emph{there
exists at least one interpretation compatible with quantum mechanics
in which particle number is undefined}. Thus, given that systems in
states like \eqref{e:CoherentState} are predicted by QM and can
indeed be reproduced in the laboratory, we are going to propose
below a formalism in order to incorporate physical systems in such
states in a set theoretical framework.

\section{Preliminaries and primitive
symbols}\label{s:preliminaries}

We will work with a variant of Zermelo-Frenkel ($ZF$) set theory
\cite{BrignoleDaCosta} with physical things ($PTs$). We will denote
this theory by $ZF^{\ast}$. The underlying logic of $ZF^{\ast}$ is
the classical first order predicate calculus with equality
(identity). The primitive symbols of $ZF^{\ast}$ are the following

\begin{itemize}

\item Those of classical first order predicate
calculus using only
identity and the membership symbol ``$\in$"

\item the unary predicate symbol
``$\mathcal{C}()\ldots$" (such that
``$\mathcal{C}(x)$" reads ``$x$ is a set")

\item a binary predicate symbol ``$\sqsubset$"
whose meaning
will be clear below, when we give the general
mereological axioms
used in our framework

\end{itemize}

\nd $ZF^{\ast}$ concerns sets and PTs (which are not sets), and so,
it is involved with a kind of mereology. PTs are meant to represent
physical objects. Depending on the particular interpretation of our
framework, PTs may represent fields, particles, strings or any
collection of physical objects whose interpretation is compatible
with the ontology intended for our framework. In particular, we will
consider the system represented by a state such as the one of
\eqref{e:CoherentState}, \emph{as formed by an undefined number of
photons}.

Definitions of formulas, sentences (formulas without free
variables), bound variables, free variables, etc., are the standard
ones. As usual, we write ``$\exists_{\mathcal{C}} x (F(x))$" instead
of ``$\exists x(\mathcal{C}(x)\wedge F(x))$" and
``$\forall_{\mathcal{C}}x(F(x))$" instead of ``$\forall
x(C(x)\longrightarrow F(x))$".

$ZF^{\ast}$ possesses axioms of two different kinds: the ones
concerned with sets and the ones concerned with PTs. Let us begin by
listing the set theoretical axioms.

\subsection{Set theoretical axioms}\label{s:SetTheoryAxioms}

The following postulates constitute an adaptation of those of Zermelo-Frenkel set theory (see \cite{BrignoleDaCosta} for details).

\begin{axiom}[Extensionality]
$$(\forall_{\mathcal{C}}x)(\forall_{\mathcal{C}}
y)((\forall z)(z\in x
\longleftrightarrow z\in y)\longrightarrow x=y)$$
\end{axiom}

\begin{axiom}[Union]
$$(\forall x)(\forall
y)(\exists_{\mathcal{C}}t)(\forall z)(z\in
t\longleftrightarrow (z\in x\vee z\in y))$$
\end{axiom}

\begin{axiom}[Power set]
$$(\forall_{\mathcal{C}}x)(\exists_{\mathcal{C}}
y)(\forall_{\mathcal{C}}t)(t\in
y\longleftrightarrow t\subseteq x)$$
\end{axiom}

\noindent If $F(x)$ is a formula, $x$, $y$ and $z$ are distinct variables and $y$ does not occur free in $F(x)$,
we have

\begin{axiom}[Separation]
$$(\forall_{\mathcal{C}}
z)(\exists_{\mathcal{C}}y)(\forall x)(x\in
y\longleftrightarrow F(x)\wedge x\in z)$$
\end{axiom}

\begin{axiom}[Empty set]
$$(\exists_{\mathcal{C}}t)(\forall x)(x\notin t)$$
\end{axiom}

\begin{axiom}[Amalgamation]\label{a:union}
$$(\forall_{\mathcal{C}} x)((\forall y)(y\in
x\longrightarrow \mathcal{C}(y))\longrightarrow
(\exists_{\mathcal{C}}z)(\forall t)(t\in
z\longleftrightarrow (\exists v)(v\in x\wedge t\in
v)))$$
\end{axiom}

\noindent If $F(x,y)$ is a formula and the variables satisfy evident conditions we have:

\begin{axiom}[Replacement]
$$(\forall x)(\exists
!y)(F(x,y))\longrightarrow(\forall_{\mathcal{C}}
u)(\exists_ {\mathcal{C}}v)(\forall y)(y\in
v\longleftrightarrow(\exists x)(x\in u\wedge
F(x,y)))$$
\end{axiom}

\begin{axiom}[Infinity]
$$(\exists_{\mathcal{C}}z)(\emptyset\in
z\wedge(\forall x)(x\in z\longrightarrow
x\cup\{x\}\in z))$$
\end{axiom}

\begin{axiom}[Choice]\label{a:Choice}
$$(\forall_{\mathcal{C}}x)\{(\forall y)(y\in
x\longrightarrow\mathcal{C}(y))\wedge(\forall
y)(\forall z)(y\in
x\wedge z\in x\longrightarrow(y\cap
z=\emptyset\wedge
y\neq\emptyset))$$
$$\longrightarrow(\exists_{\mathcal{C}}u)(\forall
y)(\exists v)(y\in
x\longrightarrow(y\cap u=\{v\}))\}$$
\end{axiom}

\begin{axiom}[Foundation]
$$(\forall_{\mathcal{C}} x)(x\neq\emptyset\wedge (\forall
y)(y\in
x\longrightarrow\mathcal{C}
(y)))\longrightarrow(\exists z)(z\in x\wedge z\cap
x=\emptyset)$$
\end{axiom}

\subsection{Axioms for PTs}\label{s:ThingsAxioms}

Now we list the axioms for PTs. We will use small Greek letters for variables restricted to PTs. Informally, the symbol ``$\sqsubset$" will express the ``being part of" relation. Thus, ``$\alpha\sqsubset\beta$" means that ``$\alpha$ and $\beta$ are PTs and $\alpha$ is a part of $\beta$". We start with
some preliminary definitions.

\begin{definition}[Disjointness]\label{
d:disjointPTs}
$$\alpha|\beta:=\neg\exists\gamma(\gamma\sqsubset\alpha\wedge\gamma\sqsubset\beta)$$
\end{definition}

\noindent $\alpha|\beta$ is interpreted as ``$\alpha$ and $\beta$
are PTs which share no part in common"; a possible definition of
indistinguishability could be given as follows (though we will not
use it in this work)

\begin{definition}[Indiscernibility]\label{
d:IndiscerniblePTs}
$$\alpha\equiv\beta:=\alpha\sqsubset\beta\wedge\beta\sqsubset\alpha$$
\end{definition}

\noindent $\alpha\equiv\beta$ means that $\alpha$
and $\beta$ are indistinguishable, in the sense that they cannot
be discerned by any physical means.

\begin{definition}[PT]\label{d:PTsDefinition}
$$T(x):=\neg\mathcal{C}(x)$$
\end{definition}

\noindent $T(x)$ reads ``$x$ is not a set", and thus, it is a PT.

\begin{definition}[Sum of
parts]\label{d:SumOfParts}
$$\mathcal{S}(x,\alpha):=\mathcal{C}(x)\wedge\forall y(y\in x\longrightarrow T(y))\longrightarrow\forall\gamma(\gamma|\alpha\longleftrightarrow\forall\beta(\beta\in x\longrightarrow\beta|\gamma))$$
\end{definition}

\nd The explanation of $\mathcal{S}(x,\alpha)$ is that if $x$ is a
set such that all its elements are PTs, then for every $\gamma$
which satisfies being disjoint to $\alpha$, then it will also be
disjoint to any element $\beta$ in $x$ and viceversa. Intuitively,
the only PT $\alpha$ which has this property is the \emph{physical
sum} of all the PTs belonging to $x$.

We now formulate a general axiomatic for PTs. These axioms may
encompass a general class of entities, ranging from field quanta to
non relativistic particles. But it is important to remark that all
these entities need more specific axioms in order to be fully
characterized; we are concentrating here in their general
mereological porperties.

We start by stating that every thing is a part of
itself

\begin{axiom}\label{a:partofitself}
$$(\forall\alpha)(\alpha\sqsubset\alpha)$$
\end{axiom}

\nd It is reasonable to assume transitivity of the
relationship
``$\sqsubset$"

\begin{axiom}\label{a:transitivity}
$$(\forall\alpha)(\forall\beta)(\forall\gamma)(\alpha\sqsubset\beta\wedge\beta\sqsubset\gamma\longrightarrow\alpha\sqsubset\gamma)
$$
\end{axiom}

\nd We will postulate that there exists the sum of any non empty set of PTs

\begin{axiom}\label{a:ExistenceSetOfParts}
$$(\forall
x)(\exists\alpha)(\mathcal{S}(x,\alpha))$$
\end{axiom}

\section{Things with undefined number of parts}\label{s:AsolutionToTheProblem}

We will use the following notation

\begin{definition}\label{d:TheSetSatisfyingF}
$$\exists\{x\,|\,F(x)\}:=(\exists y)(\forall
x)(x\in y\longleftrightarrow
F(x))$$
\end{definition}

\nd and the following definition will allow us to present a possible solution to the problem posed in Section
\ref{s:Introduction}

\begin{definition}\label{d:CantorianThing}
$$Cant(\alpha):=\exists\{\beta\,|\,
\beta\sqsubset\alpha\}$$
\end{definition}

\nd If $Cant(\alpha)$ we will say that \emph{$\alpha$ is
Cantorian}\footnote{We use ``Cantorian'' in analogy with the system
NF of Quine \cite{QuineFLPV,Rosser}. But this should not lead to any
confusion: the analogy is not too deep.}. The above definition says
that if a PT $\alpha$ is cantorian, then, all parts of $\alpha$ form
a set (and vice versa). Thus, it is possible to assign a cardinal to
any Cantorian thing $\alpha$ by assigning a cardinal number to its
set of parts in the usual way (using choice axiom \ref{a:Choice}).
Notice that it is straightforward to show that if $\alpha$ is
Cantorian, then there exists \emph{only one} set satisfying the
equality of definition \ref{d:CantorianThing}.

For any $x$ such that $\mathcal{C}(x)$, denote
$\sharp(x)$ the cardinal assigned in the usual way using the $ZF$
axiomatic (and we can use it for sets, because the axiomatic of
$ZF^{\ast}$ includes that of $ZF$). Thus we define

\begin{definition}
If $Cant(\alpha)$, let $z$ be the only set
satisfying the equality of definition \ref{d:CantorianThing}. Then we define the
\emph{cardinal} of $\alpha$ (abreviated as
$\sharp(\alpha)$) as $$\sharp(\alpha):=\sharp(z)$$
\end{definition}

\nd Any PT $\alpha$ will be cantorian or not. If
$\alpha$ is not Cantorian (i.e., if $\neg(Cant(\alpha))$), then,
there is no means for ensuring that its parts form a set using the
above axioms. Because of this, there is no way in which we can
assign to $\alpha$ a cardinal using $ZF$ axioms, and from this point
of view, it is reasonable to interpret a non Cantorian PT as
having no cardinal. In this way, we find that the axiomatic framework
presented in this work is useful to represent PTs with undefined
number of constituents as the ones presented in section
\ref{s:UndefinedNumberOverview}. But once this general solution is
presented, new problems may be posed. We list them
below:

\begin{enumerate}

\item We provided a general axiomatic for PTs. But
it is clear that each theory and spatio-temporal setting will have its
own and characteristic ontological features implying its
particular axiomatic. Which should be the specific axioms for
non relativistic quantum mechanics and relativistic quantum
mechanics respectively?

\item How to represent a physical thing which is
in a superposition state like the one represented by equation
\eqref{e:CoherentState}?

\item How to represent a physical superposition in
general?

\item Related to (1) and (2), how to represent
entanglement?

\end{enumerate}

In this work, we presented a possible solution for question 2.
Systems formed of an undefined particle number are represented by
non-cantorian things. But -up to now- our formalism does not
distinguishes the state $a_{1}|n\rangle + a_{2}|m\rangle$ from
$a'_{1}|n\rangle + a'_{2}|m\rangle$ (with $a'_{1}\neq a_{1}$ or
$a'_{2}\neq a_{2}$). In future works, we will essay possible
solutions for the problems posed above.

Notwithstanding, something can be said about
superpositions using non-cantorian sets right now (thus providing a
partial answer to question 3). The following construction, shows
that non-cantorian sets possess unexpected properties, which are
capable to yield non-standard mathematics and can represent
physical situations at the same time. Suppose that $\alpha$ is such that
$\neg(Cant(\alpha))$. Then, given a formula
$F(x)$, it is impossible --with the above axioms-- to grant the existence
of the set

\begin{equation}
\alpha_{F}=\{\beta\sqsubset \alpha\,|\,F(\beta)\}
\end{equation}

\noindent The separation axiom \emph{cannot be applied}, because the
parts of $\alpha$ do not conform necessarily a set! But in a
standard set theory (like $ZF$), ``properties" are usually expressed
as the membership to given set. For example, if we want to state
that the number $4$ is even, we can express this by the formula
$4\in\{x\in\mathbb{N}\,|\,\exists y(x=2\times y\wedge
y\in\mathbb{N})\}$. But if we want to interpret our formula $F(x)$
as representing a physical property in $ZF^{\ast}$ (defined by
extension as the set of all PTs possessing that property), we will
face a problem. We cannot grant the existence of the set formed by
the parts of $\alpha$ possessing the property defined by $F(x)$.
This is a direct consequence of $\neg(Cant(\alpha))$. This situation
could be interpreted as follows: ``if $\alpha$ is not Cantorian, we
cannot assert that its parts possess the property defined by $F(x)$
or that they do not possess it". This fact, does not constitutes a
real problem for our framework, but an unexpected advantage:
\textit{this kind of undetermination in the possession of a property
can be interpreted as being in a superposition state}. Indeed, a key
feature of a quantum mechanical superposition is the lack of meaning
in asserting or denying the possession of a given property.

When we face a superposition ---say, in a system of spin
$\frac{1}{2}$--- such as $\frac{1}{\sqrt{2}}
(|\uparrow\rangle+|\downarrow\rangle)$, we are not capable of
asserting that the system has spin up nor spin down: this is a key
aspect of superpositions, captured by our framework. Thus, our
framework is also capable of giving a precise logical meaning to
superpositions (at least of a special kind). In order to make thinks
clearer, think of $\alpha$ as formed by the photons of a state of
the electro magnetic field such as \eqref{e:CoherentState}. As it is
a superposition in particle number, its energy is also undefined,
and thus, the set of photons possessing a definite energy value will
inherit the non-Cantorianity of $\alpha$.

\vskip3mm

Taking into account the above discussions, it would be interesting to provide a definition of
what should be considered classical and quantum PTs within our
framework. We give definitions below trying to capture such notions.

\begin{definition}[Irreducible
Part]\label{d:IrreduciblePart}
$$\mathcal{I}(\alpha,
\beta):=\alpha\sqsubset\beta\wedge(\forall
\gamma)(\gamma\sqsubset\alpha\longrightarrow\gamma
\equiv\alpha)$$
\end{definition}

\nd $\mathcal{I}(\alpha,\beta)$ will be interpreted as ``$\alpha$ is
an irreducible part of $\beta$", and this means that $\alpha$ is a part of $\beta$ and that any part of $\alpha$ will
be indistinguishable of $\alpha$ itself. It is straightforward to show that if $\alpha$ is cantorian, then there exists
the set of all irreducible parts (hint: use separation). We
remark that this set may be the empty set. Now we will define the
important notions of \emph{classical part} and \emph{quantum part} with
respect to a well formed formula $F(x)$. If $\alpha$ is a PT and
$F(x)$ is a formula, we define

\begin{definition}
$Cant_{F}(\alpha):=\exists\{\beta\sqsubset\alpha\,
|\,F(\beta)\}$
\end{definition}

\noindent If $Cant_{F}(\alpha)$ we will say that \emph{$\alpha$ has a cantorian subset of parts satisfying $F(x)$}. If
$\neg Cant_{F}(\alpha)$, we will interpret this as: ``some parts of $\alpha$ are in a superposition state with respect to the property
$F(x)$". Thus, given a formula $F(x)$, we will say that

\begin{definition}[Quantum
Part]\label{d:QuantumPart}
$$\mathcal{QP}_{F}(\alpha):=\neg
Cant_{F}(\alpha)$$
\end{definition}

\noindent and interpret this as: ``$\alpha$ is quantal with respect
to $F(x)$".

\begin{definition}[Classical
Part]\label{d:ClassicalPart}
$$\mathcal{CP}_{F}(\alpha):=Cant_{F}(\alpha)$$
\end{definition}

\noindent and interpret this as: ``$\alpha$ is
classical with respect to $F(x)$".



We conclude this Section by adding a list of
general remarks which could be useful to consider in further
developments of a mereology involving quantum entities.

\begin{enumerate}

\item As remarked above, different axioms could be
added to the above framework in order to capture different
kinds of PTs. The specific form of these axioms will depend on the
particular physical theory but also -and strongly- on the
interpretation of that theory.

\item It should be clear that the spatio-temporal
setting in which the theory is developed (v.g., Galilean space time
for non-relativistic quantum mechanics and Minkowski
space-time for QFT) have a crucial influence in the mereological
properties of the corresponding physical objects. This implies that,
in order to develop a more specific framework, axioms
containing specific space time notions should be added to the axiomatic
presented in this work.

\item We may represent a general physical system
as a triplet $<P,M,S>$, where $P$ is a set representing PTs,
$M$ is the corresponding space-time differential manifold of
the theory and $S$ is a mathematical structure involving mathematical
objects, some of which are built with the help of $M$. For example,
non-relativistic quantum mechanics may be represented as a set, endowed with Galilean
manifold and the axiomatic of von Neumann written in the
mathematical language of functional analysis. A unitary transformation will
thus be a mathematical concept linked to the space-time
notion of Galilean symmetry transformation. It is important to remark
that the explicit inclusion of the space time manifold, while
necessary for experimental verification of the theory, does not
implies necessarily that the entities involved has well
defined spatio-temporal properties, as is the case in the
orthodox interpretation of QM.

\item It is easy to show that if in our system there are Cantorian
sets, then, the totality for PTs will not be a set.

\item If one wants to quit identity of our system (in order to consider indistinguishable objects as in \cite{fyk}),
it suffices replace identity ``$=$" for a new symbol ``$\equiv$",
postulating that it is an equivalence relation with extra conditions
(chosen in a suitable way in order to capture the desired physical
features).

\end{enumerate}

\noindent In future works, we will address these questions by
developing a new system, namely $Z^{\ast\ast}$, capable of
incorporating all these features, and thus, providing a complete
quantum mereology. The development of a quantum mereology is still
an open problem, and the formal framework presented here is a
concrete step in this direction. In particular, the formal approach
to quantum features of our system is not present in previous
mereological discussions (as for example, in
\cite{Simons,Darby,Borghini-Lando}).

\section{Conclusions}\label{s:FinalConclusions}

\nd In this work we presented a solution for what can be considered
a generalization of the Manin's problem, namely, the problem of
\emph{incorporating in a set theoretical framework the quantum
mechanical notion of undefined particle number}. Furthermore,
\emph{our system recovers the interesting feature of possessing
undefined properties representing quantum superpositions}. Although
our proposal is a valid solution for the problems posed in
\cite{Holik-Thesis,Holik-Thesis2,Domenech-Holik,
Holik-NoNameNonumber, Domenech-Holik-deRonde}, a lot of questions
arise and remain unsolved. In particular, it would be interesting to
search for other axiomatic systems capturing quantum entanglement.

By incorporating these quantum features, \emph{our framework is a
concrete step for the development of a rigorous quantum mereology}.
This is an important issue for those philosophers interested in the
development of any ontology which takes quantum mechanics as a
fundamental theory.

Of course, many other constructions could be envisaged, and they may
depend on the particular interpretation of the quantum formalism.
For example, it would be interesting to look for the specific
implications that the spatio-temporal setting has for the
mereological axiomatic capturing the properties of the physical
systems of different theories. In particular, a quantum relativistic
and non-relativistic mereology is lacking, and we think that the
development of set theoretical frameworks like the one presented in
this work could be useful for that purpose.

The characterization of undefined particle number and more general
quantum superpositions presented in this work, could be used in
different --and perhaps, more sophisticated-- frameworks. We note
that the proposed logical system presented in this paper can be used
as a basis for all non relativistic quantum mechanics; we shall
discuss this question in a forthcoming paper.

Another interesting question to look at would be
that of the implications for mathematics of systems like the
one presented here. How would it be a
mathematics not based on our every day concepts, but on quantum
mechanics? Such a question was partially answered \cite{fyk}, but our system opens a new door to such a research program. In
particular, the system presented above, constitutes a novel
example of non-standard mathematics, which gives a precise logical meaning
to the --up to now-- intuitive notion of what physicists mean by
``undefined particle number".

\vskip1truecm

\noindent {\bf Acknowledgements}

\end{document}